\documentclass[10pt,letterpaper]{article}

\usepackage{tabu}

\usepackage[top=0.85in,left=2.75in,footskip=0.75in]{geometry}

\usepackage{changepage}
\usepackage[utf8x]{inputenc}
\usepackage{textcomp,marvosym}
\usepackage{amsmath,amssymb}
\usepackage{cite}
\usepackage{nameref,hyperref}
\usepackage[right]{lineno}
\usepackage{microtype}
\DisableLigatures[f]{encoding = *, family = * }

\raggedright
\usepackage[aboveskip=1pt,labelfont=bf,labelsep=period,justification=raggedright,singlelinecheck=off]{caption}

\makeatletter
\renewcommand{\@biblabel}[1]{\quad#1.}
\makeatother
\date{}

\usepackage{lastpage,fancyhdr,graphicx}
\usepackage{epstopdf}
\fancyheadoffset[L]{2.25in}
\fancyfootoffset[L]{2.25in}
\usepackage{array}

\begin{document}
\vspace*{0.2in}

\begin{flushleft}
{\Large
\textbf\newline{Structure in scientific networks: towards predictions of research dynamism} 
}
\newline
\\
Benjamin W. Stewart\textsuperscript{1},
Andy Rivas\textsuperscript{2},
Luat T. Vuong\textsuperscript{3,4},
\\
\bigskip
\textbf{1} Expository Writing Program, New York University, New York, NY, USA
\\
\textbf{2} Conductor, Inc. New York, NY, USA
\\
\textbf{3} Department of Physics, Queens College of the City University of New York, Flushing, NY, USA
\\
\textbf{4} Department of Physics, The Graduate Center of the City University of New York, New York, NY, USA
\\
\bigskip

%
%





* Luat.Vuong@QC.cuny.edu (LTV), BWS5@nyu.edu (BWS), arivas@conductor.com (AR)

\end{flushleft}
\section*{Abstract}

Certain areas of scientific research flourish while others lose advocates and attention. We are interested in whether structural patterns within citation networks correspond to the growth or decline of the research areas to which those networks belong. We focus on three topic areas within optical physics as a set of cases; those areas have developed along different trajectories: one continues to expand rapidly; another is on the wane after an earlier peak; the final area has re-emerged after a short waning period. These three areas have substantial overlaps in the types of equipment they use and general methodology; at the same time, their citation networks are largely independent of each other. For each of our three areas, we map the citation networks of the top-100 most-cited papers, published pre-1999. In order to quantify the structures of the selected articles' citation networks, we use a modified version of weak tie theory in tandem with entropy measures. Although the fortunes of a given research area are most obviously the result of accumulated innovations and impasses, our preliminary study provides evidence that these citation networks' emergent structures reflect those developments and may shape evolving conversations in the scholarly literature.

\section*{Introduction}

\par Sociologists, geographers, and systems theorists have long analyzed how network structures shape the forces that pass across them \cite{bettencourt,burt,granovetter,putnam2001bowling,weng2012competition}. Much of this work traces its roots back to the “strength of weak ties” theory, which Mark Granovetter developed as a way of linking "micro-level interactions to macro-level patterns" \cite{granovetter}. This theory asserts that weak social ties (friends-of-friends rather than close friends) are more likely to provide access to novel information than strong ties. Ronald Burt’s structural hole theory \cite{burt} builds on Granovetter’s work via its assertion that the strength of a tie is less important than its position in the structure. Burt places emphasis on the importance of non-redundant ties, which exist in cases where a social actor ({\it ego}) connects two other people who would otherwise have no tie ({\it alters}). Within that structure, ego has powerful brokerage opportunities. The insight of Burt's theory is that those opportunities emerge from a structural feature of social networks: nonredundant ties are tantamount to social power; the egos that connect such ties together play an outsized role in the processes through which information and resources pass across those links in the social network.

The question we ask of citation networks concerns the emergence and evolution of research areas: are there relations between an area's development and the shape of its citation network structures? As noted above, scholars in fields other than sociology have borrowed network theory's structural insights. For instance, Arbesman {\it},---in their work on innovation in cities---cite Granovetter to help them explain why larger cities tend to produce more innovations per capita than smaller ones \cite{arbesman2009superlinear}. Similarly, Peter Csermely, in looking at the importance of bridging structures in complex systems, identifies nodes that he refers to as ``creative elements" \cite{csermely2008}. Such nodes, Csermely argues, ``seem[] to be a widespread feature of evolving systems" and he speculates that they ``enable the survival of unprecedented challenges and play a key part in the development and evolvability of complex systems" \cite{csermely2008}. Though Csermely's scientific research examines protein-protein interaction networks, he draws on  structural hole theory to explain isomorphic phenomena in a range of situations: in ``protein interaction networks, signaling networks, social networks and ecosystems, we can also identify highly similar creative elements" \cite{csermely2008}. Burt has also linked structural holes to creativity, arguing that ``good ideas are disproportionately in the hands of people whose networks span structural holes" \cite{burt2004structural}; this assertion squares with recent findings that scholarly performance strongly correlates with access to structural holes \cite{thieme2007perspective,abbasi2012egocentric}. 

However, citation networks are strikingly different from social networks. Whereas a scholar may actively broker ideas between peers who otherwise lack ties to each other, beyond the initial act of referencing, relations among articles are inert: citation networks only evolve as scholars add references to them. In what ways then, do the insights of social network theory still apply to citation networks? Co-citation analysis has long understood citations as representations of concepts \cite{small1978cited,small1980co,braam1991mapping,small1999visualizing}. To the extent that this is so, innovative development of scientific literature---guided by disciplinary training---can be understood as brokerage between concepts (or aspects of theories) that have yet to be connected to each other. Our study looks at the structural features that are the precipitates of such combination and asks whether citation networks exhibit structural patterns that correspond to the dynamism of the research areas to which those networks belong. Of the citation networks of particular articles, we ask about the relations between two properties: the number of articles connected to the network via a single link (i.e., nodes that, relative to a large percentage of the other nodes in the network, are weak ties), and the variety of pathways that tie the network's other nodes together (network entropy). In combination, these properties enable us to characterize both the breadth and diversity of networks as they emerge over time. We refer to the precipitates of that emergence as a {\it citation ecology}---i.e., a pattern of relations among references. Such patterns are important to our study in that they allow us to draw new connections between network theory and citation analysis.

To see how this is so---and also to get a sense of the conceptual motivations behind our approach---it will help to understand a problem that emerged in the 1980s and 90's in citation analysis. Several scholars have presented excellent overviews of attempts, in those years, to establish what Wouters \cite{wouters1999citation} describes, retrospectively, as a desire for ''a definitive citation theory'' \cite{wouters1999citation,LeydesdorffCitationTheory1998,van1998matters,bornmann2008citation,moed2006citation}. Depending on who was calling for such a theory, it would have provided a framework with which to either a) coherently assess the value of citations as quantitative performance measures (the position affiliated with the Mertonian \cite{mertonDemocracy,merton1988matthew}, normative model of science \cite{cronin1981agreement,cronin1981need,cole1974social}), or b) question the validity of the normative model's assumptions (the position most strongly associated with constructivist sociology, and Science and Technology Studies \cite{cozzens1989,edge1977not,edge1979quantitative}). During this period, attempts to establish a normative theory of citation failed due both to the wide variety of citer motivations
\cite{cronin1981agreement,cronin2005hand,moravcsik1979citation,krampen2007validity}, and to the disparate methodological approaches to accounting for that variety, a problem well-summarized by Bornmann and Daniel \cite{bornmann2008citation}. Although the constructivist sociologists of science tended to see those problems as endemic to the  project of using qualitative measures as assessments of scientific quality, scientometricians responded with approaches that justified normative science on grounds in line with what Van Raan characterizes as ''bibliometric chemistry''. This approach, rather than focusing on the idiosyncratic referencing behaviors of individual scholars, focuses on the statistical distributions that emerge within fields and sub-fields \cite{van1998matters}. Wouters \cite{wouters1999citation} characterizes such strategies as generating formalized rather than paradigmatic representations of science; he notes that the former have become the dominant mode of citation analysis.

Our study constructs formalized representations of citation network structures in order to approach the citation behavior question from a different angle. We ask whether the patterns that emerge from aggregations of citation behaviors signal dynamics operating within the paradigmatic dimension of particular research areas. As such, we are interested in citation networks not for what they tell us about the articles they reference; rather, we use those cited articles as focal points through which to access and analyze the structural character of the networks connected to them. Those networks represent ongoing, scholarly conversations that establish avenues that scientists may add to as part of their future scholarship. Our methods characterize the structure of those conversations, an approach distinct from co-citation and co-word analysis\cite{small1974structure,small1978cited,osareh1996bibliometrics,he2002mining,bertin2016linguistic}, and from those co-authorship analyses that have employed network theory in order to understand the kinds of  positions and relations that are most advantageous to scholarly production and recognition \cite{thieme2007perspective,abbasi2012egocentric}. Whereas those studies seek to either, (a) isolate the significant networks of concepts that have emerged from scholarly conversations, or (b) focus on the advantages of particular structural positions within the social network, we look at citation network structure to see if patterns of referencing behavior correlate with the dynamism of areas of scientific research.

We build and study citation networks of the leading papers in three research areas. In order to characterize those networks' citation ecologies, we focus on the relationships between two properties whose mutual growth---on multi-year time horizons---appears to be vital to continued network growth. Those two qualities are, first, the network's friends-of-friends (in this case, references to papers not already in the local citation network), and second, the diversity of network pathways as measured by Shannon entropy and normalized Shannon entropy. We see these properties as being particularly important components of our networks' citation ecologies due to their potential to move in opposite directions---i.e., spikes in the intensity of one property lead to drops in the normalized measure of the other. On the other hand, when these properties increase or decrease at more gradual rates, they often move in the same direction. By focusing on these measures, we begin to explore the extent to which patterns within a citation network correspond to the more general dynamics of the research area to which it belongs.

\section*{Constructing citation networks in three areas of research}
We focus on three topic areas within optics: plasmonics\footnote{Plasmons are coherent excitation of electrons on a surface or interface, generally a metal.} (topic search: 'plasmon*'), the photorefractive effect\footnote{Photorefractive effect: self-induced changes in refractive index due to photon-photon interaction in a material.} (topic search: `photorefractiv* NOT keratect*'\footnote{Refractive keratectomy is unrelated to the physics principle of the photorefractive effect, the innovation channel isolated in this study.}, and Brillouin scattering\footnote{Brillouin scattering refers to the photon response to self-induced distortions in atomic  lattice.} (topic search: `brillouin scatter*'). Fig. 1 shows each topic's popularity since 1975: plasmonics continues to grow at a striking rate, with over 11,000 journal articles published within 2015; on the opposite trajectory, photorefractives are on the wane; finally, research related to Brillouin scattering, the smallest field surveyed here, re-emerged after having slowed in the late-1980s and early-90s. As of 2015, plasmon is the largest topic area (with over 62,000 papers) and its articles have attracted the most citations. Photorefractive is the second largest area, with about 8,000 papers. Brillouin scattering is the smallest area with approximately 6,000 papers. We refer to the three fields as plasmon ({\it growing}), Brillouin scattering ({\it re-emerging}) and photorefractive ({\it waning}).   

\begin{figure}[t]
\centering
\includegraphics[width=11.4cm,]{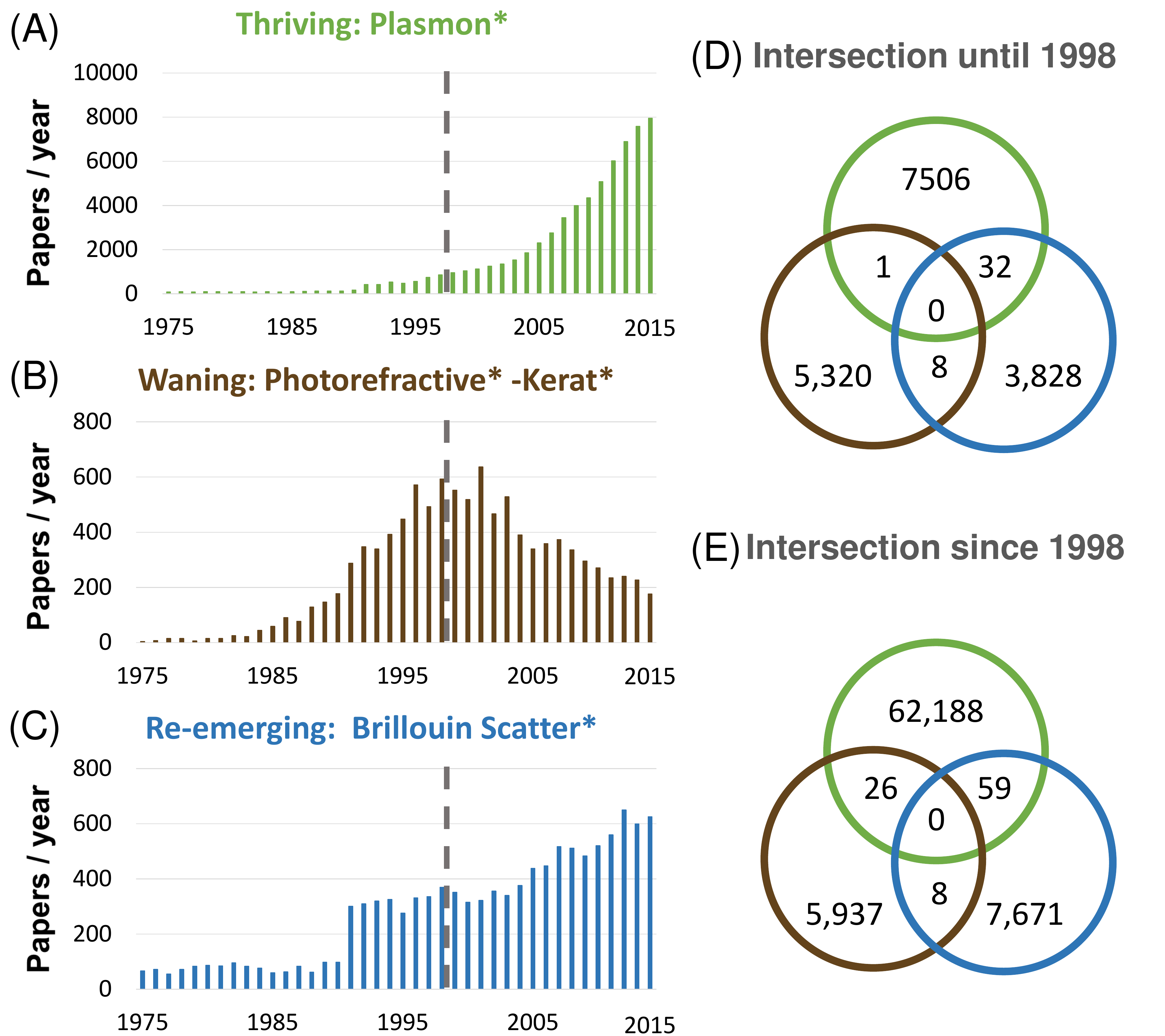}
\caption{Three topics within the field of optics that exhibit separate dynamics in research activity.  The numbers of papers published per year from 1975 through 2015 show topics (A) `plasmon*', an area that is growing (B) `photorefract* NOT keratect*', whose topic association is waning, and (C) and `Brillouin scatter*', whose topic is re-emerging. Venn diagrams with the number of papers published in the `plasmon*' (green), `photorefract* NOT keratect*' (brown), and `Brillouin scatter*' (blue) fields (D) before and (E) after 1998. Note that the scale of the plasmon topic area shown in (A) is approximately an order of magnitude larger than the other two areas.}\label{AnalRecon0}
\end{figure}

In the 1970s these three topic areas emerged out of the widespread availability of laboratory lasers. As such, there is substantial overlap in terms of the equipment and laboratory setup necessary to conduct research. All three topics involve both theorists and experimentalists; prior to 1998, all three were dominated by optical physicists. Research operates on a moderate scale for facilities---i.e., experiments typically do not require large collaborative networks or large accelerator facilities (e.g., those at CERN); while all are highly dependent on the development of materials (as the basis for experiments), the ease of production and the costs of the technologies vary widely; results from all three areas have led to the development of commercial technologies, although those applications lie in different domains. Among the three, there is almost no overlap in research interests or citation networks; as of 2014, only 60 papers connected photorefractives to plasmons, only 30 papers connected Brillouin Scattering to plasmons, and only 8 papers connected photorefractives to Brillouin scattering. The separation between these areas allows us to compare their network structures and also to consider the dynamics of each one in relative isolation from the others. 

For each topic area, we identify the 100-most cited articles with publication dates before 1999. Although 1998 is a somewhat arbitrary endpoint, it marks a time during which the use of scholarly databases---as well as the development of online, often open-source, scholarly publishing---was expanding \cite{VandeSompel2000}. Limiting the publication period of top-cited articles to pre-1999 allows us to reduce the effects of any catalyzation of publicity stemming from immediate online notification and access. Given their citational significance, we refer to each of our top-cited papers as a {\it parent} and create a database of each parent's {\it family}. Each of the 300 network families range in size from 800 to 20000 members and includes {\it grandparents} (papers referenced by the parent), {\it descendants} (papers that reference the parent), and the descendants' other references, some of which may already exist within the network as primary ties to the parent. We refer to those articles that cite the parent directly as {\it endogenous nodes}; by contrast, we refer to those articles that are two steps removed from the parent as {\it exogenous nodes} (i.e., nodes that, to a parent, are friends of friends). Although exogenous nodes have a range of potential relations to the scholarly conversation established by the parent, they don't explicitly participate in it.

Our scraping process sweeps the citation network in three passes in order to produce a citation network. It gathers, first, the references for each parent (the grandparents); then the papers that cite the parents (descendants); finally, for each of these descendants, we collect their references to other papers (thus, no paper is located more than two vertices away from the parent). With this data, we construct a directed network for each parent wherein each node represents a paper and each edge a reference link between two nodes $u\rightarrow v$ indicates that paper $u$ references paper $v$. Although we construct the networks out of directed pathways, we analyze them as undirected networks. Primarily, this approach stems from our {\it disinterest} in the citational capital with which citation networks endow significant articles\footnote{Another reason for this approach has to do with the way that citation indices are constructed in the direction opposite to the scholarly literature's emergence: references look backwards; citations look forward in time \cite{wouters1999citation}.}. We focus on how the commitments of a given parent network are structured, particularly in terms of the relations between its exogenous and endogenous nodes; structurally, the direction of commitment is not relevant. 

\section*{High citation reach and entropy marks growing and re-emerging research}
\begin{figure}[!ht]
 \includegraphics[width=5.25in]{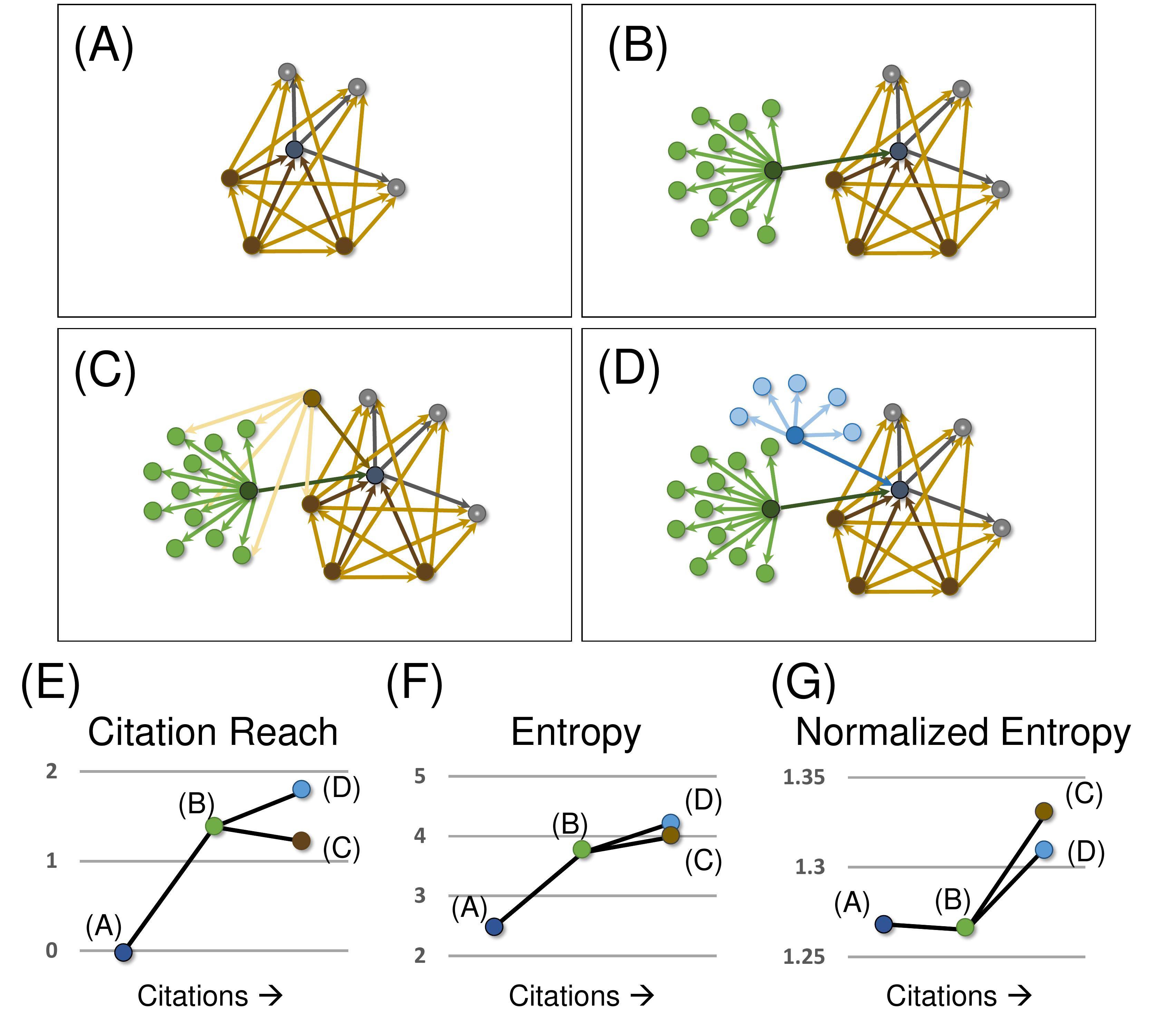}\centering
 \caption[Caption for LOF]{The structure of parent-paper citation networks exhibit structural changes based on the citation practices of its descendants.\footnotemark (A) represents a parent-paper network containing the basic network of 3 grandparents and 3 descendants; that network's structure changes significantly when (B) a new descendant introduces a cluster of 11 exogenous nodes. A subsequent descendant may cite the network's exogenous nodes, as in (C), or, alternatively and by contrast, add exogenous nodes to the parent citation network, as in (D).}
\label{NetworkPlots}
\end{figure}

As each parent network grows, its shape emerges in part as a function of the descendants' reference distribution---i.e., the number of those descendant references that are endogenous and the number that are exogenous, as illustrated in Fig. \ref{NetworkPlots}. As discussed above, exogenous nodes are analogous to the social network model of friends-of-friends. One of our observations is that the parent papers of the growing [plasmon] and re-emerging [Brillouin] fields both exhibit network structures that more closely resemble Figs. \ref{NetworkPlots}(B) and (D) rather than structures of Figs. \ref{NetworkPlots}(A) and (C), even though the re-emerging field is an order of magnitude smaller than the growing area. Figures \ref{NetworkPlots}(B,D) each have a larger network than (A,C) for the same number of citations; for (B,D), the new descendants (new nodes) that enter the network by citing the parent also cite endogenous rather than exogenous nodes. The networks of the parent papers in the growing and re-emerging areas have networks that exhibit higher entropy per citation (entropy grows quickly in a network whose new citations add large numbers of exogenous nodes). The growing and re-emerging areas' networks also have larger numbers of nodes per citation than the waning network; we refer to the nodes-per-citation measure as {\it citation reach}.

\footnotetext{Note that, in reality, it would be rare to have a citation network as small and as insular as Fig. 2 (A); that network is an ideal type designed to contrast with the other three networks. Likewise, (C) and (D) highlight extremes of network potential.}

To measure of the number of nonredundant ties in the parent networks, we calculate those networks' citation reach. This concept is related to, though less precise than, reach efficiency, which measures ''how much (nonredundant) secondary contact'' a given node has ''for each unit of primary contact'' \cite{scott2011sage}. In a similar manner, the citation reach, $R$, quantifies the number of exogenous nodes in the network per parent and descendant:
\begin{equation}
R = \frac{N-C-G-1}{C+G+1}
\end{equation}
where $N$ is the number of nodes in the network, $C$ is the number of citations of the parent, $G$ is the number of references by the parent or grandparents. As suggested above, this measure should not be understood as a property of the parent paper; rather, it's a property of the ecology of the network tied to the parent. While there exist other formulations for calculating the embedded relationship between players in a network \cite{backstrom2014romantic}, this study draws uses more basic measures to characterize simpler networks. Here, entropy and citation reach describe the parent citation networks that, containing only four levels, are not complex but can already be shown to be quite distinct. In fact, it is remarkable that such basic measures can differentiate structural patterns within these citation networks. 

\begin{center}
\begin{figure}[h!]
\includegraphics[width=5in]{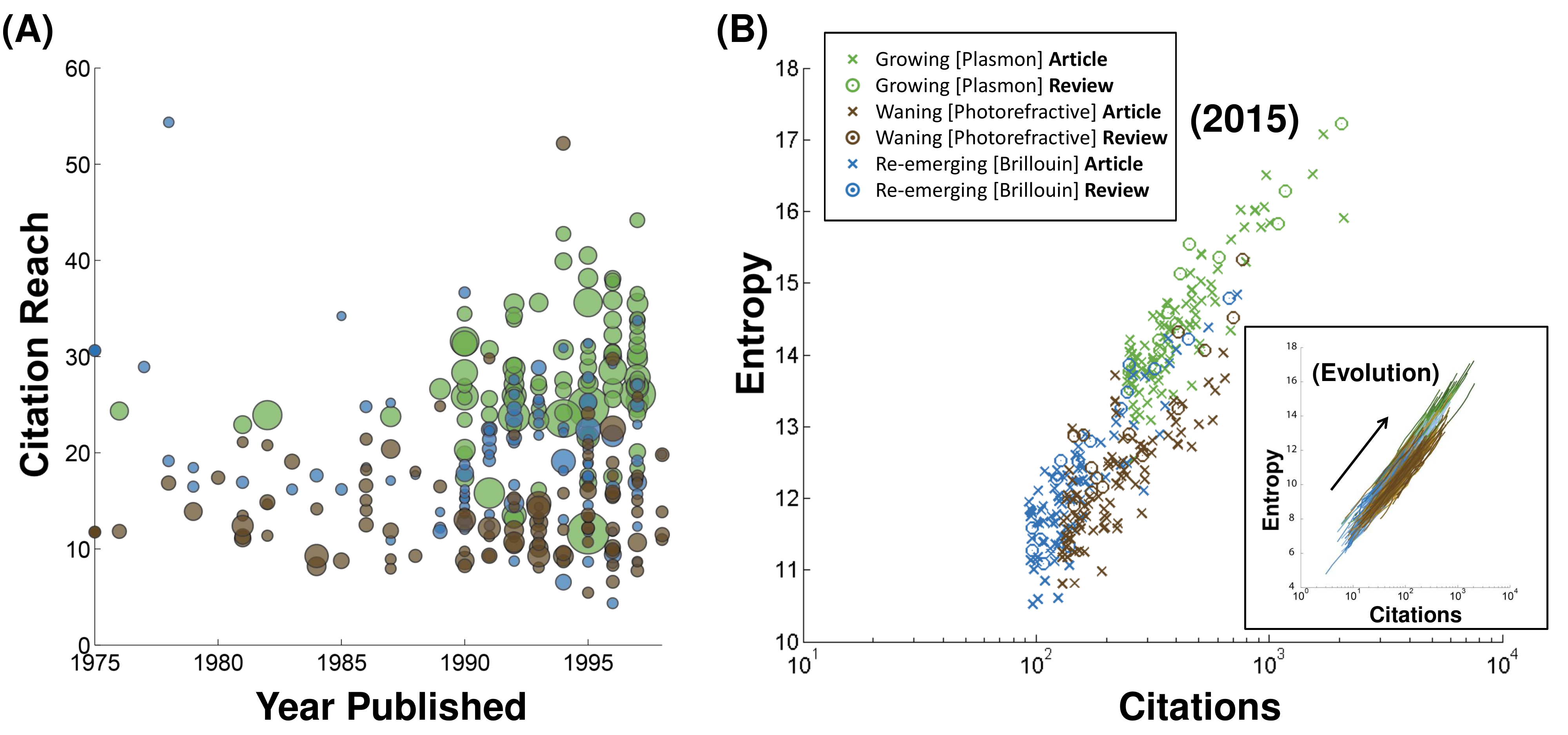}
\caption{The structure of the citation networks of the leading parent papers in each field may indicate the dynamics of the field. (A) Citation reach as a function of the year of publication for each of the 300 citation networks represented in this study, where the circle radius indicates the number of parent-paper citations. Parent-paper networks in the growing [plasmon] (green) and re-emerging [Brillouin] (blue) fields exhibit higher citation reach than those in the waning [photorefractive] (brown) field. (B) Parent-paper network entropy as a function of parent-paper citation at the end of 2015.  Parent-paper networks in the growing [plasmon] (green) and re-emerging [Brillouin] (blue) fields exhibit higher entropy per citation than those in the waning [photorefractive] (brown) field. The three selected topics occupy relatively distinct areas of the figures. No obvious differences are observed between parent papers that are research articles (X's) and those that are review papers (circles). Inset: the evolution of each parent paper's network entropy is drawn as a function of its citations over time.}
\label{AnalRecon2}
\end{figure}
\end{center}

In Figure \ref{AnalRecon2}(A), we show the citation reach $R$ of each field's leading parent papers as a function of the parent-paper's year of publication. The radius of each circle is scaled to the number of citations of the parent paper. We observe that the growing and re-emerging fields [plasmon and Brillouin] exhibit higher $R$ than the waning field [photorefractives]. The majority of the more highly-cited parent papers chosen for this study from each field were published between 1990 and 1998; however, the $R$ for the growing and re-emerging areas is relatively uniform over time in spite of this nonuniform temporal introduction of parent papers. 

Whereas $R$ measures the number of exogenous nodes per descendant, it does not describe the integration of those, and other, nodes into the network. Thus, in addition to $R$, we characterize another aspect of the parent papers' network ecologies, namely their Shannon entropy $ S = -\sum_i^N p_i \log(p_i)$ and normalized Shannon entropy $H = \frac{S}{\log{N}}$, where $N$ is the number of papers or nodes in the secondary network, and $p_i$ is the fraction of citations related to paper $i$. The values of $S$ and $H$ provide measures of network growth and interconnectedness---characterizations of the diversity of the path structures of a given citation network. 

Within citation analysis, we are unaware of projects that use Shannon entropy in this way, though Leydesdorff \cite{leydesdorff2002indicators} has used Shannon entropy to examine citation dynamics among scientific journals, and Li {\it et al.}, \cite{li2015entropy} have used it to detect communities within citation networks. Outside of citation analysis, several studies have employed Shannon entropy with motivations similar to ours: in Balch's examination of robot behavior, entropy ``provides a continuous, quantitative measure of robot team diversity" \cite{balch2000}; in Eagle's comparison of the entropy a population's communication behavior, higher communicative entropy exhibits positive correlations with higher socioeconomic rankings \cite{eagle2010}; in Demetrius's comparison of the structure of a variety of network models, ``network entropy is positively correlated with robustness" \cite{demetrius2005}. In our study, entropy characterizes the distribution of pathways among papers in a citation network.

We observe that, as measured by Shannon entropy $S$, the three topic areas are structurally different from each other. In Fig. \ref{AnalRecon2} (B), each marker (color-coded by topic area) indicates a parent paper's network entropy relative to its citations in 2015. That there is a positive correlation between these two variables is not surprising, since $S$ rises when more nodes are added to a network. What is notable, though, is that parent networks in the growing and re-emerging fields exhibit more entropy per citation than the waning field. The distinction between the waning [photorefractive] topic area and the two other areas is distinctive. The inset charts the increase in  individual paper-networks' entropy measures over time. Similar to those paper-networks' states in 2015, each topic area occupies relatively distinct portions of the graph.

\begin{figure}[!ht] \centering
 \includegraphics[width=5in]{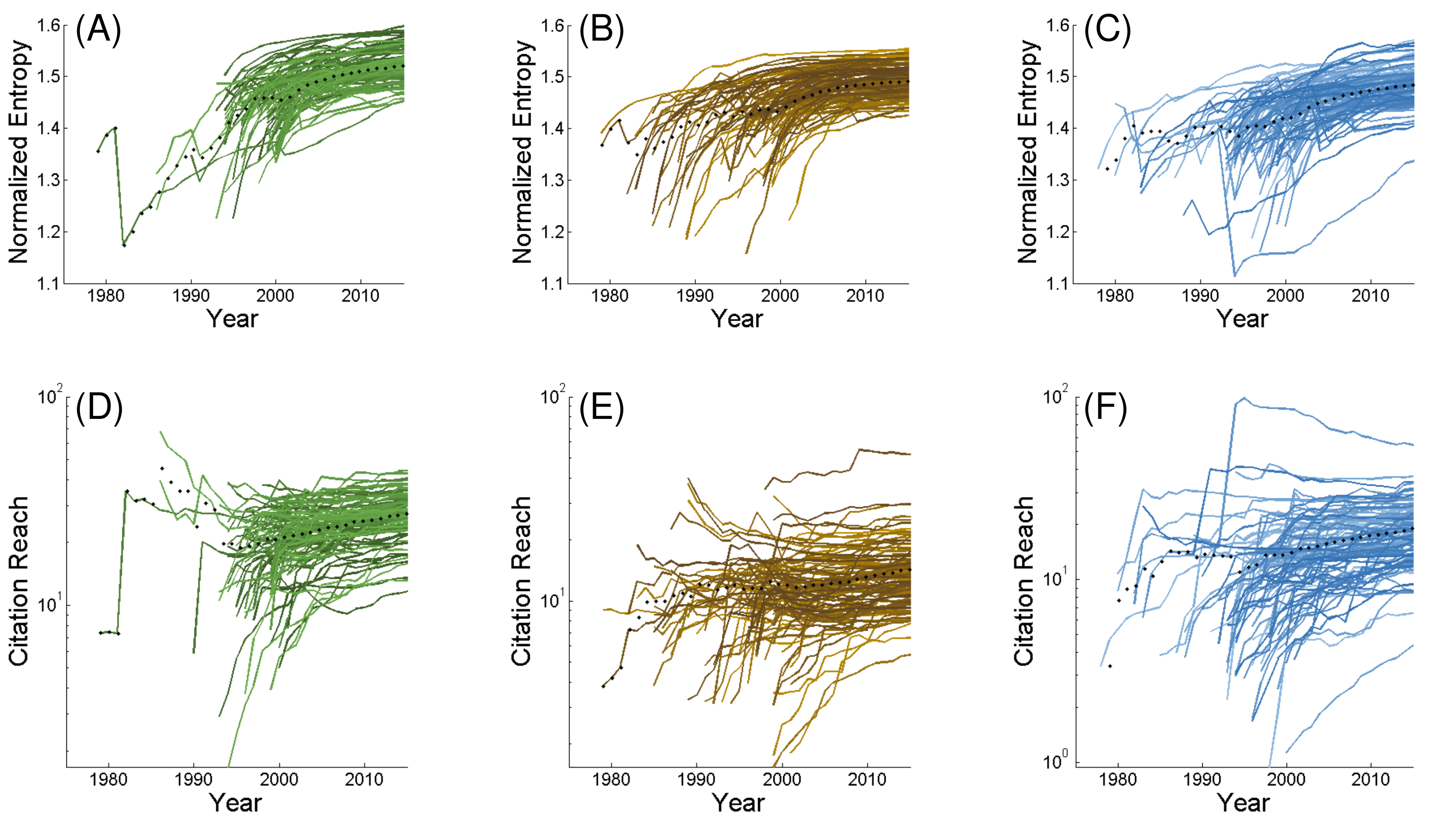}
 \caption{Punctuated drops of citation network (A)-(C) Normalized Entropy accompany punctuated jumps in (D)-(F) Citation Reach, drawn as a function of year for growing [plasmon] (green), waning [photorefractive] (brown), and re-emerging [Brillouin] (blue) fields, respectively. The black dotted lines show the average values from each field's 100 parent paper networks.}
\label{AnalRecon3}
\end{figure}

Whereas Fig. \ref{AnalRecon2} shows that citation reach and entropy per citation are higher in both growing [plasmon] and re-emerging [Brillouin] fields, Fig. \ref{AnalRecon3} illustrates that citation reach and normalized entropy can move in opposite directions: often, as part of these citation networks' growth, citation reach increases sharply along with concomitant drops in normalized entropy; many of these punctuations are followed by a gradual rise in normalized entropy (with a correlate decrease in citation reach). A contrary mathematical relationship can be deduced by assuming a linear relationship between entropy and the logarithm of citations, which is visible in Fig. \ref{AnalRecon2}(B): a jump in the parent-paper network's nodes-per-citation value denotes a large introduction of exogenous nodes, a phenomenon that is associated with increased citation reach and a decrease in normalized entropy. Although they are pervasive in all three areas, these mirrored punctuations are most clearly visible in the growing (A and D) and re-emerging (C and F) areas. This phenomenon may have implications for further research.

\section*{Reflections on the approach and future directions}
In this study, we have begun to analyze citation ecologies in relation to the dynamism of selected research areas; we have centered our analysis on simple measures that succinctly identify patterns of variation that lead to divergent outcomes. Our initial observations indicate that leading-paper citation networks carry signatures that correlate with the dynamism of the research areas to which they belong: network qualities, independent of the size of the citation network, correlate with the growth or waning of the field. These results are easily observed, though their significance is limited by our focus on only three areas within optics. Nevertheless, our results offer a place to begin to ask questions beyond those areas. Are correlations between network structure and dynamism visible across other topics of research? If so, are the correlations present in a manner analogous to our observations here, or do those properties' measures vary in different ways, perhaps depending on the type of research, or relative to the scholarly cultures conducting that research?

Even within the case study conducted here, the variations may stem from a range of factors: the social conventions that cultivate or undermine scientific norms \cite{smaldino2016natural}; the material costs of research (a factor tied both to technological history and to funders' interest); advancements or deadlocks within research lines that open up or close down future paths for exploration. Of these factors, research developments are easiest to identify and they offer ready explanations for our areas' divergent trajectories. In the growing [plasmon] area, advances in surface-enhanced Raman scattering \cite{VD77, moskovits1984, Otto92} and demonstrations of single-molecule detection \cite{Kneipp97} resulted in an explosion of biomedical applications that significantly expanded the area's interdisciplinarity. Similarly, in the 90s, work in the re-emergent area [Brillouin scattering] on stimulated Brillouin scattering dynamics \cite{BoydSBS,KurashimaSBS,SmithSBLaser1991} may have contributed to the renewed interest and a range of optical applications \cite{OkawachiSBS,LiSBLaser2012,GrattanSensors2000}. By contrast, the inflection point of the waning [photorefractive] area coincided with the disruptive development of organic photorefractive materials \cite{WangPhotorefractiveMatlRev2000, DucharmePRL1991, MoernerPolymer1994, MarderPolymerNature1997}, which may have shifted the attention from physics to chemistry and materials science\footnote{It should be noted that the photorefractive effect describes materials that are used in holography, but few papers in holography today use the word ``photorefractive''---an interesting subject in topic identification in itself.}. 

While our network measures retrospectively mark the divergences among our three areas, a question for future research is whether or not these methods can be adapted to capture research dynamics as they emerge. If so, focusing on citation ecologies could fill a gap between assessments that focus on individual scholars' publication records \cite{hirsch2005index,sidiropoulos2007generalized,moed2009new,yao2014ranking}, and those concerned with the research programs of larger-scale entities---such as institutions \cite{narin1976evaluative,moed2009new}, innovation networks \cite{AcemogluPNAS2016}, or nations \cite{leydesdorff1989science,moed1995new,guan2007china}. That is, by clustering together measures of the structures of individual networks, projects that characterize citation ecologies have the potential to clarify the links between the micro level of individual papers, and their relations to the larger, macro-structures to which they contribute.

Currently, there is a trend towards employing enormous data sets and sophisticated machine learning methods that, with a similar objective, seek to draw predictions from network structures. Our efforts aim to discern patterns of citation networks rather than social networks, and to identify research trends; we believe we are among the first to successfully correlate the structure of citation networks with the dynamics of research topics by strategic selection of comparable data sets.  This initial effort, with three disjoint research areas, does not yet provide broad statistical claims about the utility of our approach. However, this case study is novel in the way that it introduces an ecological perspective to the project of using leading papers in a field to embody the trends of the whole. Our results may indicate that, over the long term, diversity within citation networks is vital to the continued growth of topic areas. 

There are similarities between this ecological focus and some work in patent analysts---studies that may be instructive for future research. For instance, Wang {\it et al.}, \cite{WangPatenttAnalysis} have shown that strong brokerage positions in patent networks predict for patent renewal. In a slightly different vein, Yoon {\it et al.}, have used subject-action-object (SAO) analysis to bring together multiple variables in the production of dynamic patent maps that point to "hotspots" and "vacuums" in the patent landscape \cite{YoonPatentCitationScientometrics2013}. Yet though this work is closely related to ours, the conventions of academic scholarship are different from those of patent citation (which are driven by potential profit). Indeed, there is increasing evidence that the scholarly, scientific inputs that culminate in innovative patents do not necessarily correlate with successful scholarship---to such an extent that the most prominent scientific papers are negatively correlated with patent innovations \cite{GittelmanSciencePatent2003}. 

In future work, one could use this approach to studying leaders in a field, but move the theory forward by refining the citation ecology concept, particularly as it relates to the importance of novelty in research to citation networks. Novelty, as we mean it here, refers not to the emergence of a new idea, but to citations that are new to a particular network. Given their frequency, the punctuations exhibited in Fig. 4 may suggest that periodic infusions of exogenous nodes, as well as those nodes' later integration, are vital to the dynamism of citation networks. Understood in this way, those punctuations show a compelling similarity to the results of Tria {\it et al.}, \cite{tria2014dynamics}, in which the emergence of novelty corresponds to similar drops in system entropy. In between drops, normalized entropy grows as gaps in the network are filled. In citation networks, those punctuations may signal different kinds of network growth: on one hand, they may represent new contributions to the parent article's line of research; on the other hand, they may represent an emerging interdisciplinarity related to the parent article's conceptual concerns. As markers of such developments, those punctuations--in combination with analyses that account for the roles of exogenous nodes--may offer a ground for predicting the future developments in the growth and decline of topic areas.

 \section*{Methods and author contributions}
The data for this citation study have been obtained from Thomson Reuters' Web of Science. For each entry, we compile a list containing authors, source, volume, issue, title, pages, and publisher in order to merge duplicate entries\footnote{Even at this time, all citation indices contain some typos or inconsistencies, which are corrected on a case-by-case basis. We observe that, month-to-month, the number of papers pinned to a topic shifts within a percent. This results in significant data cleaning being necessary, which is the reason for the limited case study.}. After unique entries are identified, we increase the scraping efficiency and remove bias by assigning a unique parent ID to each publication. This ID is separately stored: all data analyses are carried out in a manner that is blind to the publication-specific information such as author, title, year of publication, and journal name. 

AR collected, compiled, and organized all data.  BWS provided research context and motivation.  LTV analyzed data, made figures, and managed project. BWS and LTV both wrote the manuscript.

\section*{Acknowledgments}
LTV graciously acknowledges NSF DMR 115-1783 for broader-impact support and Prof. Alex Gaeta, who seeded discussions on the informal processes of scientific research among his students more than a decade ago.  The authors are also grateful for helpful discussions with Prof. Larry Liebovitch and Dana Weinburg.


%

\end{document}